\documentclass
[aps,twocolumn]{revtex4}
\usepackage{epsf}

\begin{document}

\title{Universal glassy dynamics at noise-perturbed onset of chaos. A route to
ergodicity breakdown.}
\author{A. Robledo}
\address{Instituto de F\'{i}sica,\\
Universidad Nacional Aut\'{o}noma de M\'{e}xico,\\
Apartado Postal 20-364, M\'{e}xico 01000 D.F., Mexico.}

\begin{abstract}
The dynamics of iterates at the transition to chaos in one-dimensional
unimodal maps is shown to exhibit the characteristic elements of the glass
transition, e.g. two-step relaxation and aging. The properties of the
bifurcation gap induced by external noise, including a relationship between
relaxation time and entropy, are seen to be comparable to those of a
supercooled liquid above a glass transition temperature. Universal time
evolution obtained from the Feigenbaum RG transformation is expressed
analytically via $q$-exponentials, and interpreted in terms of nonextensive
statistics.
\end{abstract}


\maketitle

While at present the phenomenology of glass formation is in a
well-documented advanced stage \cite{debenedetti1} the subject
remains a prevailing and major theoretical challenge in condensed
matter physics. In experiments and numerical simulations the
transition of a liquid into a glass manifests itself as a dramatic
dynamical slowing down where the characteristic structural
relaxation time changes by many orders of magnitude in a
relatively small space of temperatures. Associated to this
process, atypical connections develop between dynamical and
thermodynamic properties, such as the so-called Adam-Gibbs
relationship between structural relaxation times and
configurational entropy \cite{debenedetti1}. These poorly
understood connections pose very intriguing questions that suggest
deep-lying, hence generic, physical circumstances which may
manifest themselves in completely different classes of systems and
therefore are capable of leading to novel but universal laws. Here
we make a case for this premise by exhibiting that glassy behavior
is in point of fact present in prototypical nonlinear maps close
to the onset of chaos.
There are clear indications that standard
phase-space mixing is not entirely fulfilled during glass forming
dynamics, i.e. upon cooling, caged molecules rearrange so slowly
that they cannot sample configurations in the available time
allowed by the process \cite{debenedetti1} \cite{debenedetti2}.
Naturally, the question arises as to whether under conditions of
ergodicity malfunction, and, as a final point, downright failure,
the Boltzmann-Gibbs (BG)\ statistical mechanics is still capable
of describing stationary states on the point of glass formation or
those representing the glass itself. The aim of this letter is to
make evident that the essential elements of glassy behavior are
all actually present within the neighborhood of the chaos
transition in simple non-linear dissipative maps. By showing this
hitherto unidentified association we put forward a minimal model
for glass dynamics that endorses the idea of universality in this
phenomenon, and at the same time provides the rare opportunity for
detailed examination of properties, such as the connection between
dynamics and statics and the departure from BG statistics.

A distinctive feature of supercooled liquids on approach to glass formation
is the development of a two-step process of relaxation, as displayed by the
time evolution of correlations e.g. the intermediate scattering function $%
F_{k}$ \cite{debenedetti1} \cite{debenedetti2}. This consists of time $t$
power-law decays towards and away from a plateau, the duration $t_{x}$ of
which diverges also as a power law of the difference $T-T_{g}$ as the
temperature $T$ decreases to a critical value $T_{g}$ \cite{debenedetti1}
\cite{debenedetti2}. This behavior is displayed by molecular dynamics
simulations \cite{kobandersen1} and successfully reproduced by mode coupling
(MC) theory \cite{gotze1}. Another important feature of the dynamic
properties of glasses is the loss of time translation invariance. This is
referred to as aging \cite{bouchaud1}, and is due to the fact that
properties of glasses depend on the procedure by which they are obtained.
The slow time decline of relaxation functions and correlations display a
scaling dependence on the ratio $t/t_{w}$ where $t_{w}$ is a waiting time.
Interestingly, a recent important development is the finding that both
two-step relaxation and aging are present in fixed-energy Hamiltonian
systems of $N$ classical $XY$ spins with homogeneous but sufficiently
long-ranged interactions. In these systems the length of the plateau
diverges with infinite size $N\rightarrow \infty $ \cite{latora1} and aging,
similar to that found for short-ranged interaction spin glasses, is observed
in the long time behavior of the autocorrelation function of system
trajectories \cite{montemurro1} \cite{pluchino1}. Remarkably, here we
demonstrate that the same two features, two-step relaxation and aging, are
displayed by simpler systems with only a few degrees of freedom, such as
one-dimensional dissipative maps at the edge of chaos in the presence of
stochastic noise, the amplitude $\sigma $ of which plays a role parallel to $%
T-T_{g}$ in the supercooled liquid or $1/N$ in the spin system. This adds to
the idea of universality for the phenomenon of glass formation.

The experimentally observed relaxation behavior of supercooled liquids is
effectively described, under simple heat capacity assumptions, by the
above-mentioned Adam-Gibbs equation, $t_{x}=A\exp (B/TS_{c})$, where the
relaxation time $t_{x}$ can be identified with the viscosity, and the
configurational entropy $S_{c}$ relates to the number of minima of the
fluid's potential energy surface (and $A$ and $B$ are constants) \cite{adam1}%
. Although at present a first principles derivation of this equation is
lacking, it offers a sensible picture of progressive reduction in the number
of configurations that the system is capable of sampling as $%
T-T_{g}\rightarrow 0$ as the origin of viscous slow-down in supercooled
liquids. As a parallel to the Adam-Gibbs formula, we show below that our
one-dimensional dissipative map model for glassy dynamics exhibits a
relationship between the plateau duration $t_{x}$, and the entropy $S_{c}$
for the state that comprises the largest number of (iterate positions) bands
allowed by the bifurcation gap - the noise-induced cutoff in the
period-doubling cascade \cite{schuster1}. This entropy is obtained from the
probability of chaotic band occupancy at position $x$.

So, our purpose here is to contrast the dynamics of a fluid near glass
formation with that of iterates in simple nonlinear maps near the edge of
chaos when subject to external noise. We chose to illustrate this by
considering the known behavior of the logistic map under these conditions
\cite{schuster1}. The idea in mind is that the course of action that leads
to glass formation is one in which the system is driven gradually into a
nonergodic state by reducing its ability to pass through phase-space-filling
configurational regions until it is only possible to go across a
(multi)fractal subset of phase space. This situation is emulated in the
logistic map with additive external noise, $x_{t+1}=f_{\mu }(x_{t})=1-\mu
x_{t}^{2}+\xi _{t}\sigma $,$\;-1\leq x_{t}\leq 1$, $0\leq \mu \leq 2$, where
$\xi _{t}$ is Gaussian-distributed with average $\left\langle \xi _{t}\xi
_{t^{\prime }}\right\rangle =\delta _{t.t^{\prime }}$, and $\sigma $
measures the noise intensity \cite{schuster1} \cite{crutchfield1}. As is
well known \cite{schuster1} \cite{beck1}, in the absence of noise $\sigma =0$
the Feigenbaum attractor at $\mu =\mu _{c}(0)=1.40115...$ is the
accumulation point of both the period doubling and the chaotic band
splitting sequences of transitions and it marks the threshold between
periodic and chaotic orbits. The locations of period doublings, at $\mu =\mu
_{n}<\mu _{c}(0)$, and band splittings, at $\mu =\widehat{\mu }_{n}>\mu
_{c}(0)$, obey, for large $n$, the power laws $\mu _{n}-\mu _{c}(0)\sim
\delta ^{-n}$ and $\mu _{c}(0)-\widehat{\mu }_{n}\sim \delta ^{-n}$, where $%
\delta =0.46692...$ is one of the two Feigenbaum's universal constants. The
2nd, $\alpha =2.50290...$ measures the power-law period-doubling spreading
of iterate positions. All the trajectories with $\mu _{c}(0)$ and initial
condition $-1\leq x_{in}\leq 1$ fall, after a (power-law) transient, into
the attractor set of positions with fractal dimension $d_{f}=0.5338...$.
Therefore, these trajectories represent nonergodic states, as $t\rightarrow
\infty $ only a Cantor set of positions is accessible within the entire
phase space $-1\leq x\leq 1$. For $\sigma >0$ the noise fluctuations smear
the sharp features of the periodic attractors as these broaden into bands
similar to those in the chaotic attractors, but there is still a sharp
transition to chaos at $\mu _{c}(\sigma )$ where the Lyapunov exponent
changes sign. The period doubling of bands ends at a finite value $%
2^{N(\sigma )}$ as the edge of chaos transition is approached and then
decreases in reverse fashion at the other side of the transition. The
broadening of orbits with periods or bands of number smaller than $%
2^{N(\sigma )}$ and the removal of orbits of periods or bands of number
larger than $2^{N(\sigma )}$ in the infinite cascades introduces a
bifurcation gap with scaling features \cite{schuster1} \cite{crutchfield1}
that we shall use below. When $\sigma >0$ the trajectories visit
sequentially a set of $2^{n}$ disjoint bands or segments leading to a cycle,
but the behavior inside each band is completely chaotic. These trajectories
represent ergodic states as the accessible positions have a fractal
dimension equal to the dimension of phase space. Thus the elimination of
fluctuations in the limit $\sigma \rightarrow 0$ leads to an ergodic to
nonergodic transition in the map and we contrast its properties with those
known for the molecular arrest occurring in a liquid as $T\rightarrow T_{g}$%
. Since the map clearly differs from a Hamiltonian model for a liquid in
that it does not take into consideration its molecular nature we may gain
information regarding universality in the processes of glass formation.

The main points in the following analysis are: 1) The quasi-stationary
trajectories followed by iterates at $\mu _{c}(\sigma )$ are obtained via
the fixed-point map solution $g(x)$ and the first noise perturbation
eigenfunction $G_{\Lambda }(x)$ of the RG doubling transformation consisting
of functional composition and rescaling, ${\bf R}f(x)\equiv \alpha
f(f(x/\alpha ))$. Positions for time subsequences within these trajectories
can be expressed analytically in terms of the $q$-exponential function $\exp
_{q}(x)\equiv [1-(q-1)x]^{1/1-q}$. 2) The two-step relaxation occurring when
$\sigma \rightarrow 0$ is determined in terms of the bifurcation gap
properties, in particular, the plateau duration is given by the power law $%
t_{x}(\sigma )\sim \sigma ^{r-1}$ where $r\simeq $ $0.6332$ or $r-1\simeq
-0.3668$. 3) The map equivalent of the Adam-Gibbs law is obtained as a
power-law relation $t_{x}\sim $ $S_{c}^{-\zeta }$, $\zeta =(1-r)/r\simeq
0.5792$, between $t_{x}(\sigma )$ and the entropy $S_{c}(\sigma )$
associated to the noise broadening of chaotic bands. 4) The trajectories at $%
\mu _{c}(\sigma \rightarrow 0)$ are shown to obey a scaling property,
characteristic of aging in glassy dynamics, of the form $%
x_{t+t_{w}}=h(t_{w})h($ $t/t_{w})$ where $t_{w}$ is a waiting time.

The dynamics of iterates for the logistic map at the onset of
chaos $\mu _{c}(0)$ has recently been analyzed in detail
\cite{baldovin1}. It was found that the trajectory with initial
condition $x_{in}=0$ (see Fig. 1) maps out the Feigenbaum
attractor in such a way that (the absolute values of) succeeding
(time-shifted $\tau =t+1$) positions $x_{\tau }$ form subsequences
with a common power-law decay of the form $\tau ^{-1/1-q}$ with
$q=1-\ln 2/\ln \alpha \simeq 0.24449$. That is, the {\it entire}
attractor can be decomposed into position subsequences generated
by the time subsequences $\tau =(2k+1)2^{n}$, each obtained by
running over $n=0,1,2,...$ for a fixed value of $k=0,1,2,...$
Noticeably, the positions in these subsequences can be obtained
from those belonging to the 'super-stable' periodic orbits of
lengths $2^{n}$, i.e. the $2^{n}$-cycles that contain the point
$x=0$ at $\overline{\mu }_{n}<\mu _{c}(0)$ \cite{schuster1}.
Specifically, the positions for the main subsequence $k=0$, that
constitutes the lower bound of the entire trajectory (see Fig. 1),
were identified to be
$x_{2^{n}}=d_{n}=$ $\alpha ^{-n}$, where $d_{n}\equiv \left| f_{\overline{%
\mu }_{n}}^{(2^{n-1})}(0)\right| $ is the '$n$-th diameter' defined at the $%
2^{n}$-supercycle \cite{schuster1}. The main subsequence can be expressed as
the $q$-exponential $x_{t}=\exp _{2-q}(-\lambda _{q}t)$ with $\lambda
_{q}=\ln \alpha /\ln 2$, and interestingly this analytical result for $x_{t}$
can be seen to satisfy the dynamical fixed-point relation, $h(t)=\alpha
h(h(t/\alpha ))$ with $\alpha =2^{1/(1-q)}$ \cite{baldovin1}. Further, the
sensitivity to initial conditions $\xi _{t}$ obeys the closely related form $%
\xi _{t}=\exp _{q}(\lambda _{q}t)$ \cite{baldovin1}. These properties follow
from the use of $x_{in}=0$ in the scaling relation
\begin{equation}
x_{\tau }=\left| g^{^{(\tau )}}(x_{in})\right| =\tau ^{-1/1-q}\left| g(\tau
^{1/1-q}x_{in})\right| ,  \label{trajectory1}
\end{equation}
that in turn is obtained from the $n\rightarrow \infty $ convergence of the $%
2^{n}$th map composition to $(-\alpha )^{-n}g(\alpha ^{n}x)$ with $\alpha
=2^{1/(1-q)}$. When $x_{in}=0$ one obtains in general \cite{baldovin1}
\begin{mathletters}
\begin{equation}
x_{\tau }=\left| g^{(2k+1)}(0)g^{(2^{n-1})}(0)\right| =\left|
g^{(2k+1)}(0)\right| \alpha ^{-n}.  \label{trajectory2}
\end{equation}

When the noise is turned on ($\sigma $ always small) the $2^{n}$th map
composition converges instead to $(-\alpha )^{-n}[g(\alpha ^{n}x)+\xi \sigma
\kappa ^{n}G_{\Lambda }(\alpha ^{n}x)]$, where $\kappa $ a constant whose
numerically determined \cite{crutchfield2}, \cite{shraiman1} value $\kappa
\simeq 6.619$ is well approximated by $\nu =2\sqrt{2}\alpha (1+1/\alpha
^{2})^{-1/2}$, the ratio of the intensity of successive subharmonics in the
map power spectrum \cite{shraiman1}, \cite{schuster1}. The connection
between $\kappa $ and the $\sigma $-independent $\nu $ stems from the
necessary coincidence of two ratios, that of noise levels causing
band-merging transitions for successive $2^{n}$ and $2^{n+1}$ periods and
that of spectral peaks at the corresponding parameter values $\mu _{n}$ and $%
\mu _{n+1}$ \cite{shraiman1}, \cite{schuster1}. Following the same procedure
as above we see that the orbits $x_{\tau }$ at $\mu _{c}(\sigma )$ satisfy,
in place of Eq. (\ref{trajectory1}), the relation
\end{mathletters}
\begin{equation}
x_{\tau }=\tau ^{-1/1-q}\left| g(\tau ^{1/1-q}x)+\xi \sigma \tau
^{1/1-r}G_{\Lambda }(\tau ^{1/1-q}x)\right| ,  \label{trajectory3}
\end{equation}
where $r=1-\ln 2/\ln \kappa \simeq 0.6332$. So that use of $x_{in}=0$ yields
$x_{\tau }=\tau ^{-1/1-q}\left| 1+\xi \sigma \tau ^{1/1-r}\right| $ or
\begin{equation}
x_{t}=\exp _{2-q}(-\lambda _{q}t)\left[ 1+\xi \sigma \exp _{r}(\lambda
_{r}t)\right]  \label{trajectory4}
\end{equation}
where $t=\tau -1$ and $\lambda _{r}=\ln \kappa /\ln 2$.

At each noise level $\sigma $ there is a 'crossover' or 'relaxation' time $%
t_{x}=\tau _{x}-1$ when the fluctuations start suppressing the fine
structure imprinted by the attractor on the orbits with $x_{in}=0$. This
time is given by $\tau _{x}=\sigma ^{r-1}$, the time when the fluctuation
term in the perturbation expression for $x_{\tau }$ becomes $\sigma $%
-independent and so unrestrained, i.e. $x_{\tau _{x}}=\tau
_{x}^{-1/1-q}\left| 1+\xi \right| $. Thus, there are two regimes for time
evolution at $\mu _{c}(\sigma )$. When $\tau <\tau _{x}$ the fluctuations
are smaller than the distances between adjacent subsequence positions of the
noiseless orbit at $\mu _{c}(0)$, and the iterate positions in the presence
of noise fall within small non overlapping bands each around the $\sigma =0$
position for that $\tau $. In this regime the dynamics follows in effect the
same subsequence pattern as in the noiseless case. When $\tau \sim \tau _{x}$
the width of the fluctuation-generated band visited at time $\tau _{x}=2^{N}$
matches the distance between two consecutive diameters, $d_{N}-d_{N+1}$
where $N\sim -\ln \sigma /\ln \kappa $, and this signals a cutoff in the
advance through the position subsequences. At longer times $\tau >\tau _{x}$
the orbits are unable to stick to the fine period-doubling structure of the
attractor. In this 2nd regime the iterate follows an increasingly chaotic
trajectory as bands merge progressively. This is the dynamical image -
observed along the {\it time evolution }for the orbits {\it of a single state%
} $\mu _{c}(\sigma )$ - of the static bifurcation gap first described in the
map space of position $x$ and control parameter $\mu $ \cite{crutchfield1},
\cite{crutchfield2}, \cite{shraiman1}.

In establishing parallels with glassy dynamics in supercooled liquids, it is
helpful to define an 'energy landscape' for the map as being composed by an
infinite number of 'wells' whose equal-valued minima coincide with the
points of the attractor on the interval $[-1,1]$. The widths of the wells
increase as an 'energy parameter' $U$ increases and the wells merge by pairs
at values $U_{N}$ such that within the range $U_{N+1}<U\leq U_{N}$ the
landscape is composed of a set of $2^{N}$ bands of widths $w_{m}(U)$, $m=$ $%
1,...,2^{N}$. This 'picture' of an energy landscape resembles the chaotic
band-merging cascade in the well-known $(x,\mu )$ bifurcation diagram \cite
{schuster1}. The landscape is sampled at noise level $\sigma $ by orbits
that visit points within the set of $2^{N}$ bands of widths $w_{m}(U)\sim
\sigma $, and, as we have seen, this takes place in time in the same way
that period doubling and band merging proceeds in the presence of a
bifurcation gap when the control parameter is run through the interval $%
0\leq \mu \leq 2$. That is, the trajectories starting at $x_{in}=0$
duplicate the number of visited bands at times $\tau =2^{n}$, $n=1,...,N$,
the bifurcation gap is reached at $\tau _{x}=$ $2^{N}$, after which the
orbits fall within bands that merge by pairs at times $\tau =2^{N+n}$, $%
n=1,...,N$. The sensitivity to initial conditions grows as $\xi _{t}=\exp
_{q}(\lambda _{q}t)$ ($q=1-\ln 2/\ln \alpha <1$ as above) for $t<t_{x}$, but
for $t>t_{x}$ the fluctuations dominate and $\xi _{t}$ grows exponentially
as the trajectory has become chaotic and so one anticipates an exponential $%
\xi _{t}$ (or $q=1$). We interpret this behavior to be the dynamical system
analog of the so-called $\alpha $ relaxation in supercooled fluids. The
plateau duration $t_{x}\rightarrow \infty $ as $\sigma \rightarrow 0$.
Additionally, trajectories with initial conditions $x_{in}$ not belonging to
the attractor exhibit an initial relaxation stretch towards the plateau as
the orbit falls into the attractor. This appears as the analog of the
so-called $\beta $ relaxation in supercooled liquids.

Next, we proceed to evaluate the entropy of the orbits starting at $x_{in}=0$
as they enter the bifurcation gap at $t_{x}(\sigma )$ when the maximum
number $2^{N}$ of bands allowed by the fluctuations is reached. The entropy $%
S_{c}(\mu _{c},t_{x})$ associated to the state at $\mu _{c}(\sigma )$ at
iteration time $t_{x}(\sigma )$ has the form $S_{c}(\mu _{c},t_{x})=$ $%
2^{N}\sigma s$, since at $t_{x}(\sigma )$ each of the $2^{N}$ bands
contributes with an entropy $\sigma s$ where $s=-\int_{-1}^{1}p(\xi )\ln
p(\xi )d\xi $ and where $p(\xi )$ is the distribution for the noise random
variable. In terms of $t_{x}$ (since $2^{N}=$ $1+t_{x}$ and $\sigma
=(1+t_{x})^{-1/1-r}$) one has $S_{c}(\mu _{c},t_{x})/s=(1+t_{x})^{-r/1-r}$
or, conversely,
\begin{equation}
t_{x}=(s/S_{c})^{(1-r)/r}.  \label{adamgibbs1}
\end{equation}
Since $t_{x}\simeq \sigma ^{r-1}$, $r-1\simeq -0.3668$ and $(1-r)/r\simeq
0.5792$ then $t_{x}\rightarrow \infty $ and $S_{c}\rightarrow 0$ as $\sigma
\rightarrow 0$, i.e. the relaxation time diverges as the 'landscape' entropy
vanishes. We interpret this relationship between $t_{x}$ and the entropy $%
S_{c}$ to be the dynamical system analog of the Adam-Gibbs formula for a
supercooled liquid.

Finally, we draw attention to the aging scaling property of the trajectories
$x_{t}$ at $\mu _{c}(\sigma )$. The case $\sigma =0$ is more readily
appraised because this property is, essentially, built into the same
position subsequences $x_{\tau }=\left| g^{(\tau )}(0)\right| $, $\tau
=(2k+1)2^{n}$, $k,n=0,1,...$ that we have been using all along. These
subsequences are relevant for the description of trajectories that are at
first held at a given attractor position for a waiting period of time $t_{w}$
and then released to the normal iterative procedure. We chose the holding
positions to be any of those along the top band shown in Fig. 1 for a
waiting time $t_{w}=2k+1$, $k=0,1,...$. Notice that, as shown in Fig. 1, for
the $x_{in}=0$ orbit these positions are visited at odd iteration times. The
lower-bound positions for these trajectories are given by those of the
subsequences at times $(2k+1)2^{n}$ (see Fig. 1). Writing $\tau $ as $\tau =$
$t_{w}+t$ we have that $t/t_{w}=2^{n}-1$ and $%
x_{t+t_{w}}=g^{(t_{w})}(0)g^{(t/t_{w})}(0)$ or
\begin{equation}
x_{t+t_{w}}=g^{(t_{w})}(0)\exp _{q}(-\lambda _{q}t/t_{w}).
\label{trajectory5}
\end{equation}
This property is gradually modified when noise is turned on. The presence of
a bifurcation gap limits its range of validity to total times $t_{w}+t$ $%
<t_{x}(\sigma )$ and so progressively disappears as $\sigma $ is increased.

In summary, we have shown that the dynamics of noise-perturbed logistic maps
at the chaos threshold exhibit the peculiar features of glassy dynamics in
supercooled liquids. These are: a two-step relaxation process, association
between dynamic (relaxation time) and static (configurational entropy)
properties, and an aging scaling relation for two-time functions. Our
findings clearly have a universal (in the RG sense) validity for the class
of unimodal maps studied. The occurrence of this novel analogy may not be
completely fortituous since the limit of vanishing noise amplitude $\sigma
\rightarrow 0$ (the counterpart of the limit $T-T_{g}\rightarrow 0$ in the
supercooled liquid) entails loss of ergodicity. The incidence of these
properties in such simple dynamical systems, with only a few degrees of
freedom and no reference to molecular interactions, suggests an
all-encompassing mechanism underlying the dynamics of glass formation. As
established \cite{baldovin1}, the dynamics of deterministic unimodal maps at
the edge of chaos is a {\it bona fide} example of the applicability of
nonextensive statistics. Here we have shown that this nonergodic state
corresponds to the limiting state, $\sigma \rightarrow 0$, $t_{x}\rightarrow
\infty $, for a family of small $\sigma $ noisy states with glassy
properties, that are conspicuously described for $t<t_{x}$ via the $q$%
-exponentials of the nonextensive formalism \cite{baldovin1}. The fact that
these features transform into the usual BG exponential behavior for $t>t_{x}$
provides a significant opportunity for investigating the crossover from the
ordinary BG to the nonextensive statistics in the physical circumstance of
loss of mixing and ergodic properties.

Acknowledgments. I am grateful to Piero Tartaglia for guiding me into the
concepts of glass formation and also for his kind hospitality at
Dipartamento di Fisica, Universit\'{a} degli Studi di Roma ''La Sapienza''.
I thank Fulvio Baldovin for contributing the figure. Work partially
supported by CONACyT grant P-40530-F.

\begin{figure}
\leftline{
 \epsfxsize=8.6cm
 \epsfbox{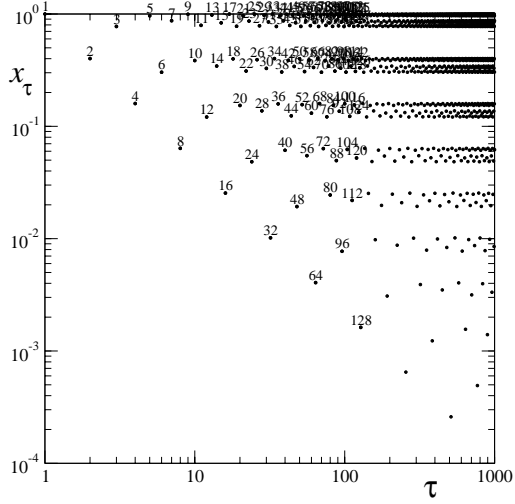}}
 \caption
  {\small Absolute values of positions in logarithmic scales of the
first $1000$ iterations $\tau $ for a trajectory of the logistic
map at the onset of chaos $\mu _{c}(0)$ \small \ with initial
condition $x_{in}=0$. The numbers correspond to iteration times.
The power-law decay of the time subsequences described in the text
can be clearly appreciated. }
\end{figure}

\end{document}